\documentclass{article}
\usepackage[T1]{fontenc}
\usepackage[latin1]{inputenc}
\usepackage{graphics}
\usepackage{setspace}
\makeatletter

\providecommand{\LyX}{L\kern-.1667em\lower.25em\hbox{Y}\kern-.125emX\@}

 \newcommand{\lyxaddress}[1]{
   \par {\raggedright #1 
   \vspace{1.4em}
   \noindent\par}
 }

\makeatother

\begin{document}

\title{The bend elastic constant in a mixture of \( 4,4^{\prime } \)-n-octyl-cyanobiphenyl
and biphenyl}

\author{Sudeshna DasGupta and Soumen Kumar Roy}

\maketitle

\lyxaddress{Department of Physics, Jadavpur University, Calcutta-700 032, India. }

\begin{abstract}
The effect of a rigid, non-polar and nonmesogenic solute biphenyl \( (C_{6}H_{5}-C_{6}H_{5}) \)
on the divergence of the bend elastic constant \( (K_{33}) \) of \( 4,4^{\prime } \)
-n-octyl-cyanobiphenyl (8CB) near the smecticA-nematic transition is reported.
The exponent decreases from about \( 1.0 \) to \( 0.67 \) as the concentration
of biphenyl increases. The temperature at which the divergence takes place is
about \( 0.5 \) to \( 1.0K \) higher than the transition temperature \( T_{AN} \).
A simple calculation based on Landau theory is presented to explore the possibilities
of the existence of a biphenyl induced tricritical point. 
\end{abstract}
Pure 8CB (\( 4,4^{\prime } \)-n-octyl-cyano biphenyl) exhibits a weakly first
order or second order smecticA-nematic transition and has been the subject of
numerous experimental studies \cite{karat77,bradshaw85,morris86,pinku94} for
over more than a decade. Despite efforts by several researchers the smecticA-nematic
transition still seems to be a major unsolved problem in equilibrium statistical
physics. A molecular field theory by McMillan \cite{mcmillan71} and a Landau
theory by de Gennes \cite{degennes72} suggest that the transition could be
both first order as well as second order, the order of transition changing at
a tricritical point. According to Halperin, Lubensky and Ma \cite{halperin74},
however, the transition can never be truly second order. 

In a recent communication \cite{dasgupta2001} we have reported a set of measurements
on a few mixtures of biphenyl (\( C_{6}H_{5}-C_{6}H_{5} \)) in 8CB. The presence
of this rigid, non-polar and nonmesogenic impurity was found to depress the
smecticA-nematic transition temperature (\( T_{AN} \)) and in the vicinity
of \( T_{AN} \) the dielectric anisotropy \( \Delta \varepsilon  \) and the
splay elastic constant \( K_{11} \) were found to exhibit anomalous behavior
for relatively higher concentrations of biphenyl. The bend elastic constant
\( K_{33} \) was found to diverge near \( T_{AN} \). Besides pure 8CB a total
of seven concentrations of biphenyl ranging from \( 0.40\% \) to \( 4.59\% \)
were used for these experiments. The findings seemed to be interesting and in
absence of any obvious explanation of the phenomena observed, we decided to
take a closer look on the behavior of the mixtures for three concentrations
of biphenyl ranging from \( 3.20\% \) to \( 4.59\% \) besides pure 8CB. These
experiments, reported in this letter, aim at finding out a more accurate behavior
of the various parameters like the dielectric anisotropy \( \Delta \varepsilon  \),
\( K_{11} \) and \( K_{33} \) in the neighborhood of \( T_{AN} \). The results
point towards the possibility of existence of a cross over from a weakly first
order or second order smecticA-nematic transition in pure 8CB and low concentration
mixtures of biphenyl towards a first order transition in mixtures richer in
biphenyl. The anomalous behavior of both \( \Delta \varepsilon  \) and \( K_{11} \)
in that they decrease as the samples are cooled towards \( T_{AN} \) is confirmed.
At the end we present a simple explanation of the shift in \( T_{AN} \) and
the changeover from a otherwise weakly first order to a first order transition
within the frame work of Landau theory. 

The experiments we have carried out involve electric field induced Freedericksz
transition. Here, temperatures could be controlled and measured with an accuracy
of \( \pm 0.1K \). From the capacitance-voltage \( (C-V) \) variation of \( 4 \)
micron sample cells with planar orientation filled with the mixtures, the dielectric
anisotropy, \( \Delta \varepsilon =\varepsilon _{\Vert }-\varepsilon _{\bot } \)
was calculated using a method which has been described in detail elsewhere \cite{dasgupta2001}.
This procedure for obtaining \( \Delta \varepsilon  \) and subsequently the
value of \( \gamma =\Delta \varepsilon /\varepsilon _{\bot } \) was first suggested
by Meyerhofer \cite{meyerhofer75}. The variation of capacitance with applied
voltage was fitted to obtain \( V_{th} \) the Freedericksz threshold voltage
and \( \kappa =K_{33}/K_{11} \). We then used the following equation to calculate
\( K_{11} \) and hence \( K_{33} \) .

\begin{equation}
\label{k11}
K_{11}=\varepsilon _{0}\Delta \varepsilon /\pi ^{2}V_{th}^{2}
\end{equation}

\begin{figure}
{\par\centering \resizebox*{0.9\textwidth}{!}{\includegraphics{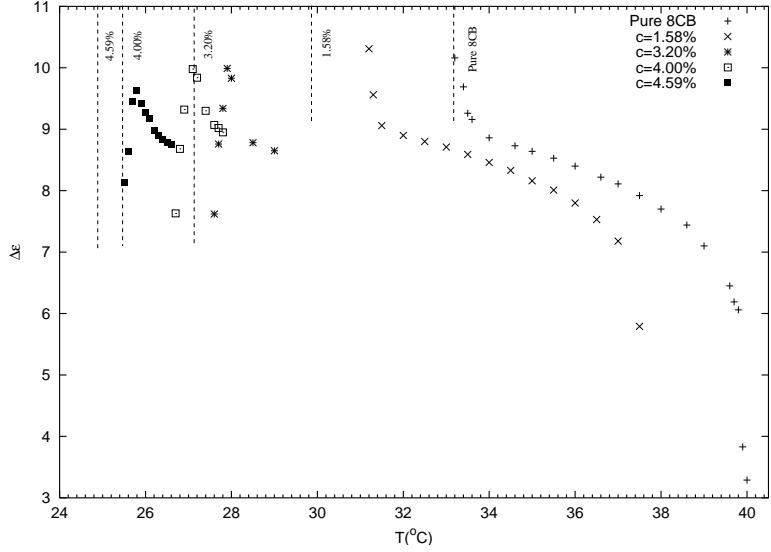}} \par}

\caption{\label{smeps}Variation of \protect\( \Delta \varepsilon \protect \) with
\protect\( T\protect \) for different concentrations of biphenyl in 8CB. The
vertical lines denote the transition temperatures \protect\( T_{AN}\protect \).
In all the figures results corresponding to c=\protect\( 1.58\%\protect \)
has been reported in \cite{dasgupta2001}.}
\end{figure}

In pure 8CB and in the low concentration mixtures of biphenyl the variation
of \( \Delta \varepsilon  \) with temperature (as \( T_{AN} \) is approached
from the nematic as well as smectic region) showed a sharp increase \cite{dasgupta2001}.
In the higher concentration mixtures studied, however, \( \Delta \varepsilon  \)
is seen to go down sharply while always remaining positive. We have noted that
the decrease in \( \Delta \varepsilon  \) as shown in Fig.\ref{smeps}, results
from a reduction of \( \varepsilon _{\Vert } \), \( \varepsilon _{\bot } \)
remaining fairly constant. A similar decrease in \( \varepsilon _{\Vert } \)
has been observed in pure \( p,p^{\prime } \) -diheptylazoxybenzene where the
\( N-S_{A} \) transition is almost second order. This has been attributed to
the presence of pretransitional effects within the nematic phase \cite{dejeu74}.

\begin{figure}
{\par\centering \resizebox*{0.9\textwidth}{!}{\includegraphics{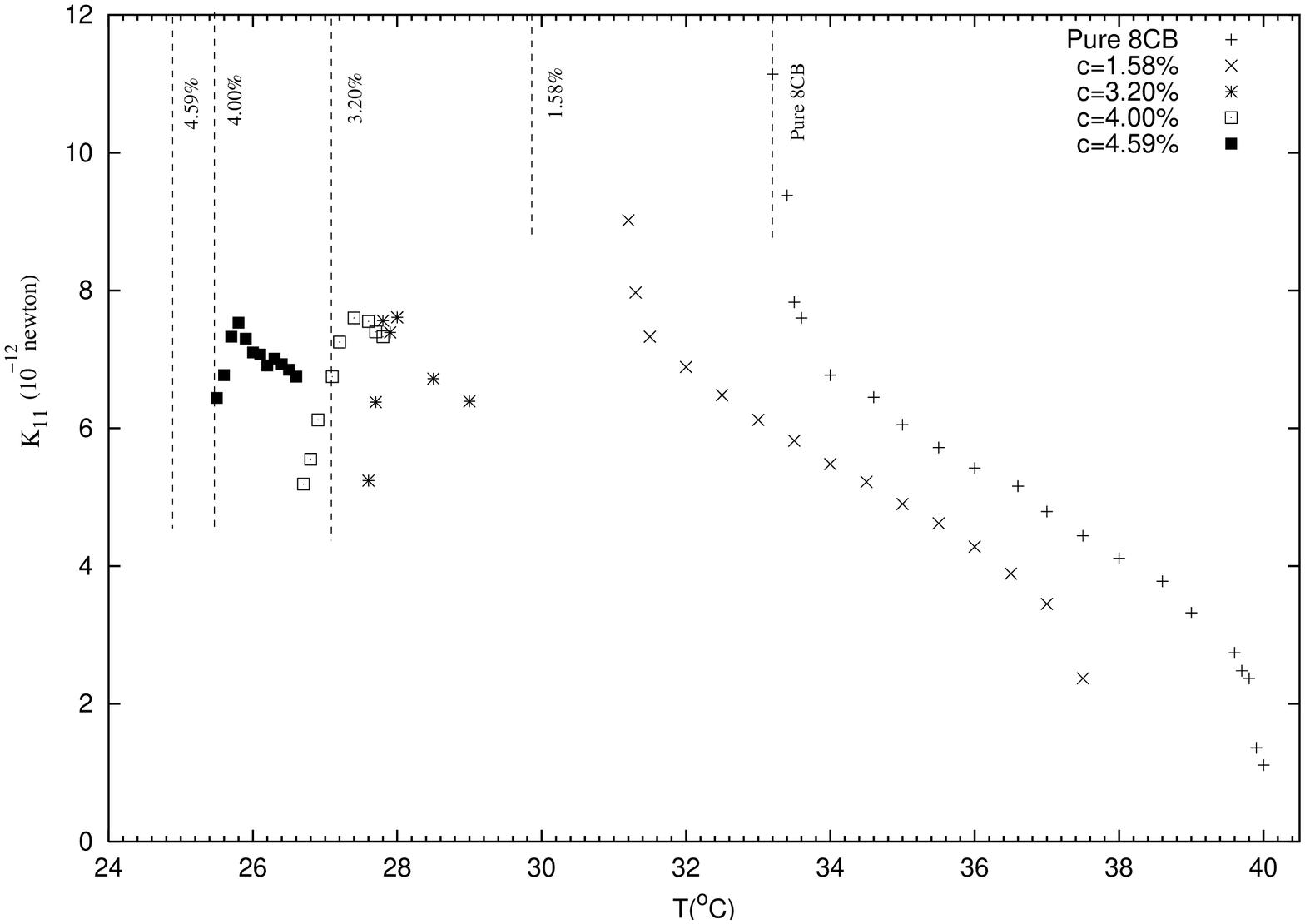}} \par}

\caption{\label{smk11}Variation of \protect\( K_{11}\protect \) with \protect\( T\protect \)
for different concentrations of biphenyl in 8CB. The vertical lines denote the
transition temperatures \protect\( T_{AN}\protect \).}
\end{figure}

The splay elastic constant \( K_{11} \) showed a variation similar to \( \Delta \varepsilon  \)
with temperature namely that it showed a sharp increase as \( T_{AN} \) was
approached in pure 8CB whereas in the mixtures it was seen to decrease sharply
with the approach of \( T_{AN} \). The variation has been shown in Fig.\ref{smk11}.

\begin{figure}
{\par\centering \resizebox*{0.9\textwidth}{!}{\includegraphics{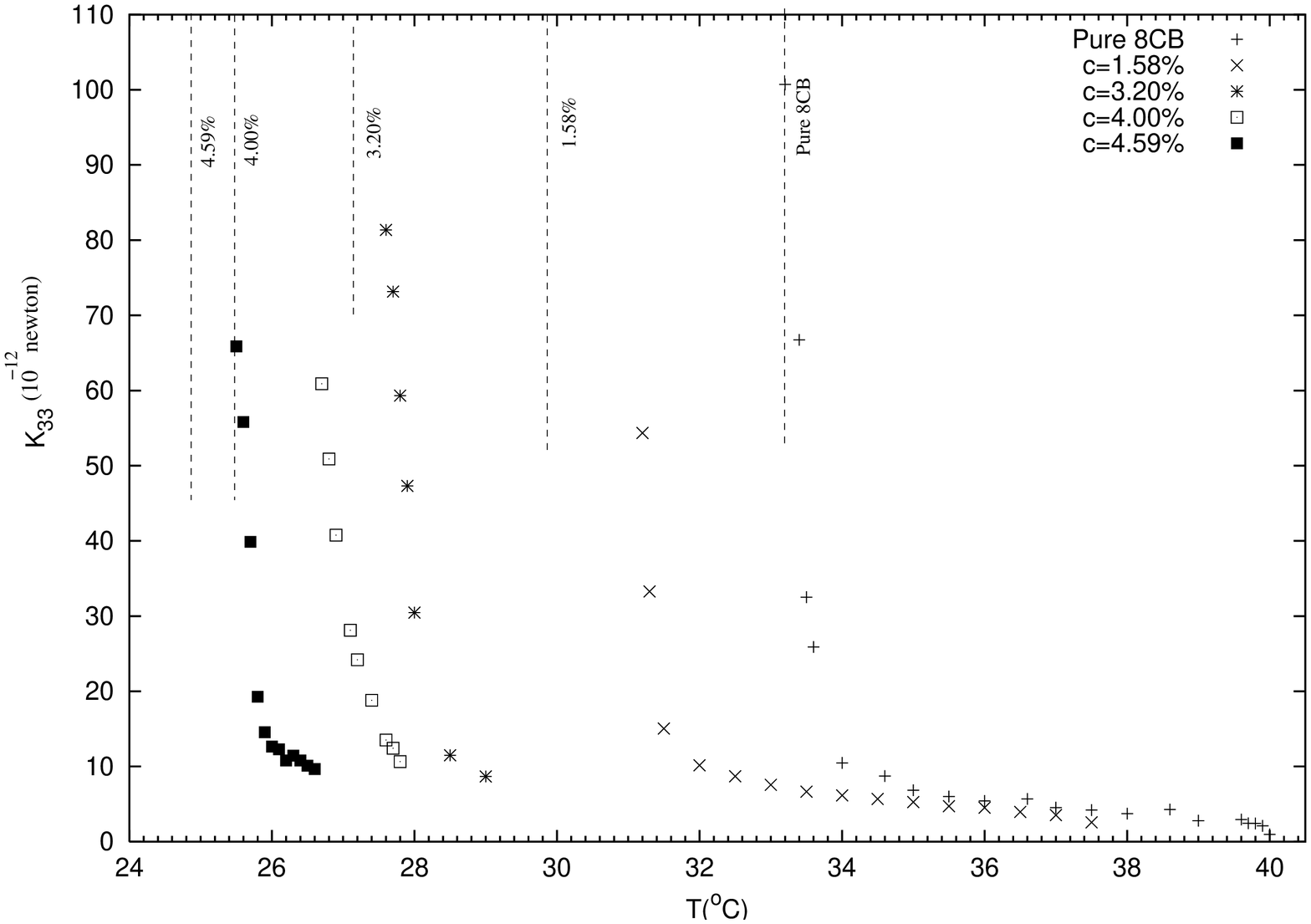}} \par}

\caption{\label{smk33}Variation of \protect\( K_{33}\protect \) with \protect\( T\protect \)
for different concentrations of biphenyl in 8CB. The vertical lines denote the
transition temperatures \protect\( T_{AN}\protect \).}
\end{figure}

The bend elastic constant \( K_{33} \) was seen to diverge as the smectic phase
was approached. The behavior was of the type \( \Delta K_{33}\propto t^{\nu } \),
where \( \Delta K_{33} \) is the difference between \( K_{33} \) and the background
nematic contribution \cite{morris86}, \( t=(T/T_{AN}^{*}-1) \) and \( \nu  \)
is the critical exponent. We found that the exponent calculated in the case
of pure 8CB was \( 0.955\pm 0.06 \) which is in agreement with the exponent
obtained by Morris et al \cite{morris86} in pure 8CB. An earlier experiment
by Davidov et al \cite{davidov79} in 8CB however yielded \( \nu =0.67 \) (which
is the same as the exponent in XY model). For the mixtures the value of the
exponent decreases as shown in Table\ref{exponent}. The temperature \( T_{AN}^{*} \)
at which \( K_{33} \) diverges is the same as \( T_{AN} \) in pure 8CB, in
mixtures it is slightly higher \( \sim  \) \( 0.5 \) to \( 1K \) than \( T_{AN} \)
. It may be recalled that anisotropic scaling laws predict that \( \Delta K_{33} \)
should vary as the correlation length \( \xi _{\Vert } \) both above and below
\( T_{AN} \) and this should result in an exponent \( \nu _{\Vert } \) which
from X-ray scattering experiments in many samples turn out to be \( \simeq 0.57-0.75 \)
\cite{degennes}. The temperature dependence of \( K_{33} \) has been shown
in Fig.\ref{smk33}.

The findings described above specially the behavior of \( K_{33} \) with temperature
near the \( S_{A}-N \) transition indicated that there might be the possibility
of a cross over from a second order or a weakly first order \( S_{A}-N \) transition
in pure 8CB towards a first order transition in mixtures. We now describe a
Landau theory calculation that may explain this.

Following de Gennes \cite{degennes}, we start by defining an order parameter
for the \( S_{A} \) phase. The \( S_{A} \) phase is characterized by a density
modulation, in a direction \( \hat{z} \) orthogonal to the layers

\begin{equation}
\label{denfluc}
\rho (r)=\rho (z)=\rho _{0}+\rho _{1}\cos (q_{s}z-\phi )+.......
\end{equation}
where \( \rho _{1} \) is the first harmonic of the density modulation and \( \phi  \)
an arbitrary phase. In a nematic \( \rho _{1}=0 \) which makes it a natural
choice for the \( S_{A} \) order parameter. 

In the vicinity of the \( N-S_{A} \) transition , the free energy per unit
volume may be expanded in powers of \( \rho _{1} \). Considering the coupling
between \( \rho _{1} \) and the nematic order parameter \( S \) it is seen
that if the alignment measured by \( S \) increases the average attractions
between molecules in a smectic layer in general increase. We define,

\begin{equation}
\label{deltaS}
\delta S=S-S_{0}(T)
\end{equation}
 where \( S_{0}(T) \) is the nematic order parameter obtained in the absence
of smectic ordering.

The coupling term between \( \rho _{1} \) and \( S \), to lowest order must
have the form

\begin{equation}
\label{nemsmcoupling}
f_{1}=-C\rho _{1}^{2}\delta S
\end{equation}
where \( C \) is a constant positive in general. The nematic free energy \( f_{N} \)
which is minimum for \( \delta S=0 \) has the form

\begin{equation}
\label{nemfreeen}
f_{N}=f_{N}(s_{0})+\frac{1}{2}\chi \delta S^{2}
\end{equation}
where \( \chi (T) \) is a response function which is large near the nematic-isotropic
transition point \( T_{NI} \) but which is small for \( T<T_{NI} \) .

We now expand the free energy in terms of the \( S_{A} \) order parameter and
include the nematic free energy and the nematic-smectic coupling term. Considering
\( x \) as the concentration of biphenyl and considering its coupling with
the nematic order parameter, the smectic order parameter and with the smectic-nematic
coupling term we have the final form of free energy

\begin{equation}
\label{freeen}
f=\frac{1}{2}r\rho _{1}^{2}+\frac{1}{4}u_{o}\rho _{1}^{4}-C\rho _{1}^{2}\delta S+f_{N}(s_{0})+\frac{1}{2}\chi \delta S^{2}+A\rho _{1}^{2}x+B\delta Sx+M\rho _{1}^{2}\delta Sx
\end{equation}
where, \( r\simeq \alpha (T-T_{0}) \), \( u_{0} \) is a positive coefficient
and \( A \) ,\( B \) and \( M \) are coupling constants. \( T_{0} \) is
a temperature at which \( r \) vanishes. With only the first two terms present
in the expression for \( f \) given by Eq.(\ref{freeen}) one would always
have a second order transition at \( T_{AN}=T_{0} \) for positive \( u_{0} \). 

Minimizing the free energy \( f \) with respect to \( \delta S \) we have,

\begin{equation}
\label{min freeen}
\delta S=\chi (C\rho _{1}^{2}-Bx-M\rho _{1}^{2}x)
\end{equation}
Substituting the value of \( \delta S \) thus obtained in \( f \), we have,

\begin{equation}
\label{final freeen}
f=\frac{1}{2}\rho _{1}^{2}r^{\prime }+\frac{1}{4}\rho _{1}^{4}u_{0}^{\prime }+f_{N}(S_{0})-\frac{1}{2}B^{2}x^{2}\chi 
\end{equation}
where we have, \begin{equation}
\label{new r}
r^{\prime }=r+2Ax+2BC\chi x
\end{equation}

\begin{equation}
\label{new u_0}
u_{0}^{\prime }=u_{0}-2C^{2}\chi +4MC\chi x
\end{equation}
to first order in \( x \). From Eq.(\ref{new r}) it is evident that the concentration
dependent terms \( 2Ax \) and \( 2BC\chi x \) depresses the transition temperature
\( T_{AN} \) provided the coupling constants \( A, \) \( B \) and \( C \)
are positive. From the concentration dependence of \( T_{AN} \) obtained in
\cite{dasgupta2001} we have \( \frac{dT_{AN}}{dx}=-1.8 \). Hence differentiating
Eq.(\ref{new r}) w.r.t \( x \) we have \( A/\alpha \approx 0.92 \) (neglecting
the term \( 2BC\chi  \) since \( \chi  \) is small near the smectic-nematic
transition).

In absence of the concentration dependent terms in Eq.(\ref{freeen}), the coefficient
of \( \rho _{1}^{4} \) is \( u=u_{0}-2C^{2}\chi  \). In the framework of this
simple mean field theory where the fluctuations are disregarded, the order of
the transition depends critically on the sign of \( u \). For \( T_{0}\sim T_{NI} \),
\( \chi (T_{0}) \) is large and \( u \) is negative. One would then require
terms in \( \rho _{1}^{6} \) in the expression for the free energy to ensure
stability and the smecticA-nematic transition would be first order. For \( T_{0} \)
significantly smaller than \( T_{NI} \), \( u>0 \) and the transition is second
order while the point \( u=0 \) gives a tricritical temperature.

In the 8CB+biphenyl system, \( T_{0} \) is significantly smaller than \( T_{NI} \)
(the difference \( T_{NI}-T_{AN} \) being \( \sim 7K \)) and the response
function \( \chi (T_{0}) \) is small. \( u \) is therefore \( \sim u_{0} \)
which is positive and a second order smecticA-nematic transition is predicted
by the theory. However the variation of the critical exponents of \( K_{33} \)
with the concentration of the biphenyl in the mixtures perhaps indicate that
there might be a cross over from a second order or a weakly first order transition
in pure 8CB towards a first order one in mixtures. This experimental evidence
suggests that the term \( 4MC\chi x \) in Eq.(\ref{new u_0}) lowers the value
\( u_{0}^{\prime } \) and this would be possible only if \( M \) is negative.
Physically this would mean an increase in the smectic-nematic coupling expressed
by the coefficient of \( \rho _{1}^{2}\delta S \) in Eq.(\ref{freeen}) from
\( C \) to (\( C-Mx \) ). It can be pointed out that at the tricritical point
\( u=0 \) and one expects that the exponent for divergence of \( K_{33} \)
here is \( \sim 0.5 \) \cite{degennes}. We also note that the magnitude of
\( M \) does not affect the shift in the transition temperature \( T_{AN} \)
as is predicted by Eq.(\ref{new r}). One may further add that the inclusion
of the last three terms in Eq.(\ref{freeen}) must produce an overall increase
in the free energy \( f \) (both \( A \) and \( B \) being positive).

To summarize, it can be stated that though the experiments and the meanfield
calculations reported in this letter do not conclusively point towards the existence
of a concentration induced tricritical temperature in the 8CB+biphenyl system,
this possibility may not be ruled out. A mixture of non-polar heptyloxypentylphenylthiolbenzoate
(\( \overline{7}S5 \)) and polar 8OCB was studied by Huster et al \cite{huster87}.
This system exhibits both a \( N-S_{A}-S_{C} \) multicritical point and a \( N-S_{A} \)
tricritical point. Because of large values of \( dT_{NA}/dx \), \( x \) being
the concentration of 8OCB the exponents suffer the so called Fisher renormalization
\cite{degennes}. The critical exponent \( \nu _{\Vert } \) associated with
the second order \( N-S_{A} \) transition in this system with \( x\simeq 0.034 \)
were observed to be \( \sim 0.58 \) (unrenormalized) and \( \sim 0.9 \) (renormalized).
The situation also resembles in many ways the change over from a second order
to a first order phase transition in metamagnets and in \( He^{4}+He^{3} \)
mixtures \cite{alben73,chaikin&lubensky}. For instance while in pure \( He^{4} \)
the normal fluid-superfluid transition is a second order lambda line, the \( He^{4}+He^{3} \)
mixture exhibits a tricritical point. The superfluid transition temperature
too is depressed in this system due to the presence of \( He^{3} \).

A more complete analysis would perhaps need to include fluctuations and the
possible effects of the presence of biphenyl on it. It must be pointed out that
we offer no explanation of the fact that the temperatures about which \( K_{33} \)
diverges is greater than \( T_{AN} \) for biphenyl concentrations> \( 1.58\% \),
a feature totally absent in pure 8CB and in mixtures with low biphenyl concentrations.

\section*{ACKNOWLEDGEMENTS}

We acknowledge valuable discussions with Dr. A. DasGupta. The research was supported
by a DST grant (SP/S2/M-20/95). SDG acknowledges the award of a fellowship.

\begin{table}
{\centering \begin{tabular}{|c|c|c|}
\hline 
Concentration (\%)&
Value of exponent (\( \nu  \))&
\( T_{AN}^{*}-T_{AN} \)\\
\hline 
\hline 
0.00&
\( 0.955\pm 0.06 \)&
\( 0.01 \)\\
\hline 
1.58&
\( 1.001\pm 0.005 \)&
\( 1.19 \)\\
\hline 
3.20&
\( 0.645\pm 0.13 \)&
\( 0.58 \)\\
\hline 
4.00&
\( 0.696\pm 0.07 \)&
\( 1.19 \)\\
\hline 
4.59&
\( 0.625\pm 0.11 \)&
\( 0.67 \)\\
\hline 
\end{tabular}\par}

\caption{\label{exponent}Values of the exponent in the mixtures studied.}
\end{table}

\bibliographystyle{unsrt}
\bibliography{lc}

\end{document}